\documentclass[fleqn,usenatbib]{mnras}

\usepackage{newtxtext,newtxmath}
\usepackage{tabularx}
\usepackage[normalem]{ulem}
\usepackage[T1]{fontenc}

\DeclareRobustCommand{\VAN}[3]{#2}
\let\VANthebibliography\thebibliography
\def\thebibliography{\DeclareRobustCommand{\VAN}[3]{##3}\VANthebibliography}

\def\farcs{$^{\prime\prime}$}

\usepackage{graphicx}	% Including figure files
\usepackage{amsmath}	% Advanced maths commands
\usepackage{xcolor}

\title[glSNe~Ia in current and future surveys]{Lensed type Ia supernovae in light of SN~Zwicky and iPTF16geu} %45 characters

\author[A. Sainz de Murieta et al.]{
Ana Sainz de Murieta$^{1}$\thanks{E-mail: ana.sainz-de-murieta@port.ac.uk},
Thomas E. Collett$^{1}$, Mark R. Magee$^{2}$, Luke Weisenbach$^{1}$,
\newauthor Coleman M. Krawczyk$^{1}$ and Wolfgang Enzi$^{1}$
\\
$^{1}$Institute of Cosmology and Gravitation, University of Portsmouth, Burnaby Road, Portsmouth, PO1 3FX, UK\\
$^{2}$Department of Physics, University of Warwick, Gibbet Hill Road, Coventry CV4 7AL, UK
}

\date{Accepted 2023 October 2. Received 2023 September 27; in original form 2023 July 24}

\pubyear{2023}

\begin{document}
\label{firstpage}
\pagerange{\pageref{firstpage}--\pageref{lastpage}}
\maketitle

\begin{abstract} 
Strong gravitationally lensed supernovae (glSNe) are a powerful probe to obtain a measure of the expansion rate of the Universe, but they are also extremely rare. To date, only two glSNe with multiple images strongly lensed by galaxies have been found, but their short time delays make them unsuitable for cosmography. Here, we simulate a realistic catalogue of lensed supernovae and study the characteristics of the population of detectable systems for different surveys. Compared to previous studies, our simulations also account for the effect of microlensing and its impact on the glSNe yields. We show that the properties of glSNe in shallow surveys (such as the Zwicky Transient Facility; ZTF) are determined by the need for large magnifications, which favours systems of four images with short time delays and low image separations. This picture is consistent with the properties of iPTF16geu and SN~Zwicky, but is not representative of the population found in deeper surveys, which are limited by the volume of the Universe that is strongly lensed. For deeper surveys, such as the Legacy Survey of Space and Time (LSST), glSNe show longer time delays and greater angular separations, and the inclusion of microlensing results in 8\% of glSNe becoming demagnified under the detection threshold. In the 10 years of the survey LSST should be able to find $\approx$ 180 systems, of which 70 will be suited for cosmography enabling a $\approx$ 1.2$\%$ precision $H_0$ measurement with LSST glSNe.
%250 words, 200 for letters
\end{abstract}

\begin{keywords}
gravitational lensing: strong -- transients:supernovae
\end{keywords}
%
%%%%%%%%%%%%%%%%%%%%%%%%%%%%%%%%%%%%%%%%%%%%%%%%%%

%%%%%%%%%%%%%%%%% BODY OF PAPER %%%%%%%%%%%%%%%%%%

\section{Introduction}
One of the biggest open questions in modern cosmology is the apparent discrepancy between early and late-time measurements of the value of the current expansion rate of the Universe, $H_0$, also known as the "Hubble tension". In the last two decades after the discovery of the accelerating expansion of the Universe through distance measurements of Type Ia supernovae  (\citealt{Riess_1998}, \citealt{Perlmutter_99}), there has been great improvement to obtain high-precision measurements of $H_0$. Local measurements based on using Type Ia supernovae (SNe~Ia) as standardisable candles provide a value of $H_0$ of $73.24 \pm 1.74$ km~s$^{-1}$~Mpc$^{-1}$ \citep{Riess_2016},
whereas Cosmic Microwave Background (CMB) measurements provide a value of $H_0=67.4\pm 0.5$ km~s$^{-1}$~Mpc$^{-1}$ \citep{planck18--20}. While the former might be subject to systematic uncertainties, the latter measurements are heavily dependent on assumptions about the cosmological model (e.g. assuming $\Lambda$CDM). This discrepancy could be an indicator for new physics beyond the standard cosmological model. Finding the reason for this tension requires alternative methods of measuring $H_0$. One powerful candidate for this is measuring the time delays between the multiple images coming from strongly lensed systems, also known as time delay cosmography.

This technique was first introduced by \cite{refsdal--64}, who suggested using gravitationally lensed supernovae to measure cosmological distances and therefore determine the cosmological parameters long before the phenomenon was first observed. Fifteen years later, the discovery of the first multiply imaged quasar \citep{walsh-79} made these sources the main subject of time delay cosmography for almost half a century. The detection of hundreds of multiply-imaged quasars and light curve monitoring led to the first robust time delay measurements (e.g. \citealt{Kundić_1997}; \citealt{Schechter_1997}). Further improvements in light curve and lens modelling during the 21st century have led to a measurement of $H_0=73.3^{+ 1.7}_{- 1.8}$ km~s$^{-1}$~Mpc$^{-1}$ (\citealt{wong--20}, \citealt{millon--20}). This result is in agreement with local measurements from SNe~Ia. 
The discovery of the first gravitationally lensed supernova in 2014 \citep{quimby--14} opened up the door to multiple advances in the field. The use of strongly lensed SNe~Ia for time delay cosmography offers several advantages over multiply-imaged quasars. Their light curves are well understood, allowing photometric data to be complemented by templates or models in order to measure time delays (e.g. \citealt{pierel19}). Alternatively, their spectra also vary with phase, making them another resource for cosmography measurements (\citealt{16geuspect}, \citealt{dttype2}). They remain bright for shorter timescales, requiring less monitoring time; and the explosion will eventually fade away after a few weeks or months, allowing for detailed reconstruction of their host galaxy and thus more accurate lens mass modelling \citep{Ding_2021}. The standardisable nature of glSNe~Ia allows for the removal of one of the main systematics in gravitational lensing: the mass-sheet degeneracy \citep{Schneider_2013}, although microlensing by stars may degrade their utility for this \citep{snmicrolensing,foxleymarrable,weisenbach}.   

Despite all these advantages, the biggest challenge in the field is that glSNe are rare and extremely hard to find \citep{magee2023}. \cite{goldstein--19} predicted that the Zwicky Transient Facility (ZTF) would detect about nine glSNe per year, out of which at least one would be a SN~Ia. In the five years of the survey however only one has been found to date: SN~Zwicky \citep{goobar--22}. Through the Bright Transient Survey (BTS; \citealt{fremling--20}), ZTF aims to spectroscopically classify all SNe brighter than $m_{g,r} \sim 19$. SN~Zwicky \citep{goobar--22} crossed this threshold and thus triggered automatic spectroscopic classification. The spectrum indicated that it was a SN~Ia at a redshift of $z = 0.35$, while its light curve appeared to be significantly brighter than expected for a SN~Ia at this redshift. It was finally found to be a quadruply-imaged SNe with total magnification of $\mu \approx 24.3$. The images appeared almost simultaneously, which is explained by the small Einstein radius of the system $\theta_{\mathrm{Ein}} \approx 0.167 ''$. This meant it was not suitable for $H_0$ measurements \citep{pierel2022lenswatch}.
SN~Zwicky shared some properties with iPTF16geu \citep{goobar--17}, a gravitationally lensed SN~Ia supernova at $z\approx$ 0.41 found with the intermediate Palomar Transient Factory \citep{iptf}, the precursor to ZTF. This supernova was similarly detected due to its extreme magnification factor of  $\gtrsim$50. The system had an Einstein radius of $\theta_{\mathrm{Ein}}\approx0.3''$, and its four images appeared also almost simultaneous, with the maximum time delay reported at 35 hours. Comparing the properties of the systems hosting SN~Zwicky and iPTF16geu with other gravitational lensing systems from both the SLACS sample \citep{auger_2009} and the SL2S sample \citep{Sonnenfeld_2013}, they both seem to come from a poorly understood population of low-mass and small image separation lensing systems that cannot be used for time delay cosmography measurements. 

Upcoming photometric surveys such as the Vera C. Rubin Observatory's Legacy Survey of Space and Time (LSST; \citealt{lsstsciencecollaboration2009lsst}) should detect more glSNe per year, but it is currently unclear whether this will allow for precise $H_0$ estimates. There have been five more confirmed multiply-imaged glSNe in cluster-scale strong lensing systems: `SN Refsdal' \citep{kelly--15}, `SN Requiem' \citep{rodney--21}, `AT 2022riv'\citep{kelly--22}, `C22' \citep{chen} and `SN H0pe' \citep{frye2023}. These systems had time delays in the order of years, which have been used to measure $H_0$. The estimate from `SN Refsdal'  is $H_0=64.8^{+4.4}_{-4.3}$ km~s$^{-1}$~Mpc$^{-1}$\citep{Kelly_2023}. The leading source of uncertainty in these measurements comes from the complexity of the cluster lens model. Galaxy-scale lenses are simpler to model, with the leading source of uncertainty in $H_0$ being the precision of the time delay measurement. This highlights the importance of searching for glSNe lensed by galaxies.

In this work we aim to answer the following questions: \textit{were iPTF16geu and SN~Zwicky outliers?}
\textit{ If not, what does this mean for the future of time delay cosmography?} 
In order to do this we simulate the population of glSNe that are discoverable based on the characteristic detection depth of photometric surveys and compare them with the properties of the iPTF16geu and SN~Zwicky. We also make predictions for the type of systems that will be detectable in upcoming surveys. We expand on previous work by \cite{goldstein--19} and \cite{wojtak19} by building light curves accounting for the effects of microlensing and using the latest version of the LSST observing strategy. We describe our simulations of lensing systems  and source light curves in Section \ref{sec:simulations}. In Section \ref{sec:discoverablesys} we provide context for current and future photometric surveys and use them to define the criteria for discoverability in our simulations.  In Section \ref{sec:results} we study the evolution of the population of detectable glSNe~Ia as a function of survey detection depth and use these results to put iPTF16geu and SN~Zwicky into context. We also obtain rate estimates for ZTF and LSST and comment on prospects for time delay measurements from LSST glSNe~Ia. In Section \ref{sec:discussion} we discuss the impact of the assumptions made in our simulations on our rate estimates and comment on the challenges that come with glSNe discovery. We summarize and conclude our results in Section \ref{sec:conclusions}. Throughout this paper we assume a flat $\Lambda$CDM cosmology with $\Omega_m=0.311$, $h=0.677$ \citep{planck18--20}. The angular diameter distances are given as $D_l$ for the distance between the observer and the lens, $D_s$ between observer and the source, and $D_{ls}$ for the distance between the lens and the source.

\section{Simulations}
\label{sec:simulations} 
In order to simulate a population of strong lenses we study the distribution of foreground lenses and their properties and background sources.
\subsection{Building a realistic population of lenses}
We follow the methods described in \cite{Collett_2015} to model the population of deflectors. We use a Singular Isothermal Ellipsoid (SIE; \citealt{sie}) model to define the density profile of our lens galaxies. This model provides a good approximation to mass profiles of elliptical galaxies. We model galaxies as ellipses with minor-to-major axis ratio $q$. The Einstein radius of an SIE is a function of the velocity dispersion of the galaxy, $\sigma_v$, and the angular diameter distances between observer, lens and source:
\begin{equation}
    \theta_{\rm{Ein}}^\mathrm{SIE} = 4 \pi \left(\frac{\sigma_v}{c}\right)^2 \frac{D_{\rm{ls}}}{D_{\rm{s}}}.
    \label{eqn:thetaein}
\end{equation}
We assume a negligible lensing contribution of matter along the line of sight. Our simple deflector model is fully described by $\sigma_v$, $q$, and the lens redshift, $z_l$. Due to translational symmetry, we place each deflector at the centre of the coordinate system and align the semi-major axis with the x-coordinate.\\ 

The velocity dispersion distribution of galaxies is given by a modified Schechter function \citep{Sheth},
\begin{equation}
\frac{dN}{dV}=dn=\phi_{*} \left(\frac{\sigma}{\sigma_*}\right)^{\alpha}\exp\left[ - \left(\frac{\sigma}{\sigma_*}\right)^{\beta} \right]\frac{\beta}{\Gamma(\alpha/\beta)}\frac{d\sigma}{\sigma},
\label{eqn:sigma}
\end{equation}
where $\Gamma$ is the gamma function and $dn$ is the differential number of lenses per unit volume. We use the parameter values derived by \cite{bernardi} from the SDSS DR6 data describing all galaxy types, $\phi_{*}=2.099\times10^{-2}$~$(h/0.7)^3$~Mpc$^{-3}$, $\sigma_*=113.78$~km~s$^{-1}$, $\alpha=0.94$, $\beta=1.85$. We draw velocity dispersions for each of the deflectors from this distribution. To obtain the redshift distribution of our lenses, we use the definition for the differential comoving volume element,
\begin{equation}
    dV=D_H\frac{(1+z_l)^2D_l^2}{E(z_l)}dz_ld\Omega,
    \label{eqn:dVol}
\end{equation}
where $D_H=c/H_0$ is the Hubble distance and $E(z_l)=\sqrt{\Omega_M(1+z_l)^3+\Omega_\Lambda}$ in our assumed cosmology. We combine Equations ~\ref{eqn:sigma} and \ref{eqn:dVol} to derive the all-sky $\left( \int d\Omega=4\pi \right)$ redshift and velocity dispersion distribution,
\begin{equation}
\frac{dN}{dz_ld\sigma}=4\pi D_H\frac{(1+z_l)^2D_l^2}{E(z_l)}\phi(\sigma).
\label{eqn:NLenses}
\end{equation}
As $\phi(\sigma)$ has no dependence on $z_l$, we can marginalize over $\sigma$ to obtain the distribution of galaxies as a function of redshift,
\begin{equation}
\frac{dN}{dz_l}=4\pi D_H\frac{(1+z_l)^2D_l^2}{E(z_l)}\int_{\sigma_{\mathrm{min}}}^{\sigma_{\mathrm{max}}}\phi(\sigma)d\sigma.
\label{eqn:pzl}
\end{equation}

We follow \cite{Collett_2015} to describe the distribution of the ellipticity for a given value of $\sigma_v$ using a Rayleigh distribution
\begin{equation}
P(1-q|s)=\frac{1-q}{s^2}\exp \Big[ - \frac{1}{2} \Big(\frac{1-q}{s}\Big)^2  \Big],
\end{equation}
with $s=(A+B\ \sigma_v)$, A\ =\ 0.38 and B\ =\ 5.7 $\times$ $10^{-4}$ km$^{-1}$ s. This distribution takes into account that more massive galaxies tend to be closer to spherical. We truncate our distribution at $q=0.2$ to exclude highly-flattened mass profiles. Sampling $z_l$, $q$ and $\sigma_v$ we can build a mock catalogue of deflectors.

\begin{figure*}
\centering
\includegraphics[width=2\columnwidth]{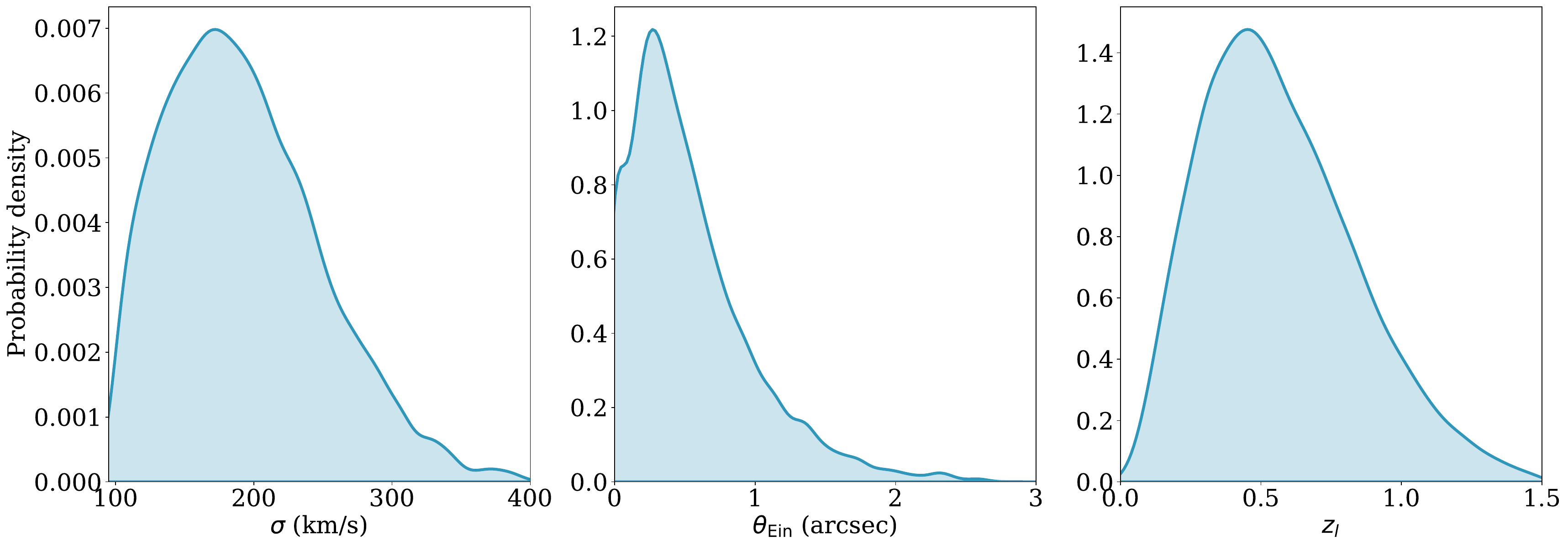}
\caption{Kernel Density Estimation of the lens parameters and Einstein radii for an ideal population of strongly lensed SNe with the assumptions provided in Section \ref{sec:simulations}.
}
\label{fig:lensdist}
\centering
\end{figure*}

Strong gravitational lensing is defined by the presence of multiple images of a source. In order for this to happen, the source must lie within the region enclosed by the radial caustic. The area in the source plane for which strong lensing will happen is known as the strong lensing cross-section, and is proportional to the square of the Einstein radius (Equation \ref{eqn:thetaein}). Thus, the probability for each of our simulated galaxies to act as a gravitational lens will depend on both the lens parameters and the redshift distribution of our source population. For the latter we use volumetric rates for SNe~Ia given by \cite{kessler--19} and convert them to redshift-dependent rates using Equation \ref{eqn:dVol}. We draw lens redshifts in the range $0\leq z_l \leq 1.5$ and velocity dispersions in the range $100\leq \sigma_v \leq 400$ km s$^{-1}$. Integrating the distribution of Einstein radii over all possible source redshifts in the range $z_l < z_s \leq 2$ we obtain the parameter distribution for galaxy-SNe~Ia lensing systems for an idealized survey, as seen in Figure \ref{fig:lensdist}. Our simulations predict that  the population of strong lenses is dominated by galaxies with $\sigma_v \approx 192.5$  km s$^{-1}$ and systems with Einstein radii of $\theta_{\mathrm{Ein}} \approx 0.53$\farcs{}. 
The median redshifts for our lenses and sources are $z_l\approx 0.57$ and $z_s\approx 1$, respectively.

\subsection{Simulating unresolved glSNe~Ia light curves}

\subsubsection{Time delay and magnification distributions}
In strong gravitational lensing, multiple images of a background point source will form at stationary points of the Fermat potential, defined as 
\begin{equation}
\tau(\theta;\beta)=\frac{1}{2}(\theta-\beta)^2-\psi(\theta),
\end{equation} 
where $\theta$ is the image position, $\beta$ is the unlensed SN position and $\psi(\theta)$ refers to the lens potential. As a result, the time delay between two images  $i$ and $j$ is given by \citep{fermattd} 
\begin{equation}
\Delta t_{ij}=\frac{D_{\Delta t}}{c}\Delta \tau_{ij},
\end{equation}
where the time delay distance (\citealt{refsdal--64}) is defined as 
\begin{equation}
D_{\Delta t}=(1+z_l)\frac{D_l D_s}{D_{ls}}.
\end{equation}
The mass profile of the lens and the supernova position in the source plane will also affect the magnification of each of its images. For the SIE, this is given by
\begin{equation}
  \mu=\frac{1}{1-2 \kappa},
\end{equation}
where $\kappa$ is the convergence at the location of the image. For the SIE this is defined as
\begin{equation}
\kappa(x_1,x_2)=\frac{\theta_{\mathrm{Ein}}}{2}\frac{\sqrt{q}}{\sqrt{x_1^2+qx_2^2}}. 
\end{equation}
We are also interested in the distribution of total magnification of the systems, which corresponds to the sum of magnifications of each of the images
\begin{equation}
\mu_{\mathrm{total}}=\sum_{i=1}^N \mu_i,
\end{equation}
where $N$ is the total number of images.
To derive probability distribution of time delays and magnifications, we assume that the the sources are uniformly distributed within the area that is strongly lensed. We sample positions uniformly in a rectangle enclosing the caustics and save those that correspond to multiple images in the lens plane (i.e are strongly lensed).
We build catalogues with distributions of time delays, magnifications, and convergence for every value of the axis ratio in the range $[0.01,0.99]$ using \texttt{lenstronomy} \footnote{\url{ https://github.com/sibirrer/lenstronomy}} software \citep{BIRRER2018189,2021JOSS....6.3283B}. 

\subsubsection{Microlensing}
In order to account for the effects of microlensing, we create microlensing maps using the inverse ray tracing method \citep{1986A&A...166...36K, 1990PhDT.......180W} for macro magnifications $\mu=\pm(2,3,5,7,10,15,20,30,40,50,60,70,80,90,100)$. We assume a constant stellar fraction $\kappa_\star = 0.5\kappa$, for all cases. The maps are convolved with a uniform disk\footnote{The size of the disk is equal to the size of a supernova that has expanded for 20 days at $10^4$ km s$^{-1}$. We take the source to have a redshift of 1.2 and the lens a redshift of 0.6. All the microlenses are taken to have a mass of $1 M_\odot$.} to approximate the expanding supernova. We then assign  histograms of the microlensing (de)magnification at peak supernova luminosity for each of the images.

\subsubsection{Unresolved light curves}
Following \cite{goldstein--19}, we make the assumption that most of our glSNe light curves will appear unresolved. We use the distribution of time delays and magnifications to simulate the unresolved glSNe light curves from the sum of the light curves for each of the images. We first use the \cite{hsiao--07} spectroscopic template to simulate unlensed SNe~Ia light curves for the $i$-band of ZTF and LSST filters in observer-frame times between $-20$ to 85\,d relative to $B$-band maximum. The peak $B$-band absolute magnitude of unlensed SNe~Ia follows a Gaussian distribution $M_B\sim \mathcal{N}(-19.3,0.2)$ mag \citep{Richardson14}.

Lensing magnifications change the flux and apparent magnitude by \citep{h0_glSN}
\begin{align}
F_X &\rightarrow \mu F_X,\\
m_X &\rightarrow m_X-2.5\textrm{log}(\mu),
\end{align}

\noindent where $\mu$ is the combined magnification from the macromodel and microlensing, and $F_X$ and $m_X$ correspond to the flux and apparent magnitude for a given filter $X$. We order the images with respect to their macromodel magnifications, so that the image with the greatest magnification is image 1. The time delays are calculated with reference to the first image. To account for the time delays of the images, we use the light curve of the image with the greatest macromodel magnification as reference. The combined unresolved glSN light curve is then given by
\begin{equation}
    F_{\mathrm{total},X}(z_s,t)=\mu_1 F_{1,X}(z_s,t)+\sum_{i=2}^N \mu_i F_{i,X}(z_s,t-\Delta t_i),
\end{equation}
where $F_X(z_s$,$t)$ is the flux derived from the spectral template, which is dependent on the SN redshift and observer-frame time, $\mu_i$ and $\Delta t_i$ are the corresponding magnifications and time delays for each of the images, and $N$ is the total number of images for each supernova. For each of our 1\,000 positions in the source plane we draw 1\,000 different values for the absolute $B$-band magnitude of the SNe and the microlensing scatter.

\section{Defining discoverable systems}
\label{sec:discoverablesys} 
In practice, not all of our supernovae will be detectable by flux-limited surveys. In this work we focus on ZTF and LSST, two of the main imaging surveys at optical wavelengths active during the periods 2018–2023 and 2024–2032, respectively. In the following subsection we briefly describe these surveys and use them to define what makes a system in our simulations discoverable.

\subsection{Optical surveys}
\label{sec:surveys}
The Zwicky Transient Facility (ZTF) is an ongoing time-domain imaging survey aimed at systematic study of the optical night sky. The ZTF camera is mounted on the 48-inch Samuel Oschin robotic telescope (P48) at the Palomar Observatory. The survey scans the entire Northern sky every two days, with its observing time divided between three major programs \citep{belm2019}: public surveys, surveys designed by members of the ZTF Collaboration, and programs selected each semester by the Caltech TAC. Most of the ZTF public surveys time is dedicated to observing the Northern Sky in the $g-$ and $r$-band every 3 days, covering a total survey area of $23\,675$ deg$^2$. As part of the ZTF Collaboration surveys there is also a slow, wide $i$-band survey observing $10\,725$ deg$^2$ of the sky with a 4 day cadence.
There are also other spectroscopic campaigns to complement the photometric surveys. One of these is the Bright Transient Survey (BTS; \citealt{fremling--20}), which aims to classify all extragalactic transients brighter than $m_{g,r}\approx 19$ mag. If a transient is found above this threshold, it will be proposed for spectroscopic follow-up and classification with the fully automated Spectral Energy Distribution Machine (SEDM; \citealt{Blagorodnova_2018}).
Our simulations require the specifications of the ZTF observing system. The median effective seeing for each of the filters is $\approx 2''$ FWHM \citep{belm2}. The 5$\sigma$ limiting magnitude for a 30 second exposure in each of the filters is $m_{\textrm{lim,g}}=20.8$, $m_{\textrm{lim,r}}=20.6$, $m_{\textrm{lim,i}}=19.9$.

\par

The Legacy Survey of Space and Time (LSST) is a 10 year optical survey of the Southern Hemisphere Sky carried out by the Vera C. Rubin Observatory set to begin operations in 2025. The "Baseline Survey Strategy" is a "wide-fast-deep" (WFD) survey covering about half of the sky every three nights  in \textit{ugrizy} filters through 15-second exposures. In addition to this, up to $10\%$ of Rubin observing time will be dedicated to other programs, including a "deep drilling" survey which will cover a smaller area of the sky at a significantly higher cadence \citep{Bianco_2022}\footnote{Specific details about the survey observing strategy are described in  \url{https://pstn-055.lsst.io/}}.  The median effective seeing in the \textit{ugrizy} bands are $(1.10, 1.03, 0.99, 0.95, 0.93, 0.92)$ arcseconds respectively. The median single-visit 5$\sigma$ depths for the WFD fields are $(23.9, 25.0, 24.7, 24.0, 23.3, 22.1 )$ in the \textit{ugrizy} bands.

To approximate these surveys and test the impact of survey depth, we run our simulations for limiting magnitudes of $m_{\mathrm{lim,i}} = 19.0, 20.5, 22.0, 24.0$. These limits approximately correspond to the BTS, the ZTF public survey, spectroscopic follow-up of LSST transients, and the LSST WFD survey.

\subsubsection{Transient recovery fraction}
\label{sec:fdet}
In reality, not all glSNe reaching a certain magnitude will be discovered. This will depend on the observing strategy of the survey, weather losses, etc. We use \texttt{simsurvey} \footnote{\url{ https://github.com/ZwickyTransientFacility/simsurvey}} \citep{feindt--19} to build survey plans and compute the parameter $f_{\mathrm{det},{m}_{\mathrm{lim}}}$, i.e. the fraction of glSNe in our simulations that would be recovered using our survey strategy. In the following subsection we describe the method used to compute this value.
For ZTF we take the one-year survey plan from \cite{feindt--19}, including observations in the $i-$band. This survey plan is characterized by constant $m_{\mathrm{lim}}=20.5$ and contains observations from the public survey, $i$-band survey, and the high cadence survey with the properties discussed in Section \ref{sec:surveys}. We use the ZTF fields and CCDs. We simulate 10\,000 glSNe using the properties of the light curves that we found would be detectable for $m_{\mathrm{lim,i}}=20.5$ mag (Section ~\ref{sec:results}) and distribute them in the area of sky covered by the survey. We check how many would be detected using our survey plan. We require two 5$\sigma$ detections in any of the bands for a glSN to be detectable. The fraction of glSNe passing these cuts over the area corresponds to the transient recovery fraction.
We follow similar steps to calculate the detectable fraction in LSST. We use the latest result from the Operations Simulator (OpSim)\footnote{The 10-years survey plan simulations \texttt{baseline\_v3.0} can be downloaded from \url{http://astro-lsst-01.astro.washington.edu:8080/}} to define our survey plan, which uses two filters at a time using a mix of uniform cadence and a half-sky rolling cadence in the "deep drilling" survey region. For the CCDs, we re-scale the ZTF corners to account for the field of view of LSST. To simulate our new lightcurves, we use light curves from our results for $m_{\textrm{lim,i}}=24$ mag. We also require two 5$\sigma$ detections in any of the bands to be detectable. 

\subsection{Sample weighting and detectability}
\label{sec:weights} 
To reduce shot noise in our results, we use importance sampling to sample lens redshifts, as their distribution contains almost no probability mass in the crucial region $z_l \lesssim 0.5$. We sample lens redshifts uniformly
\begin{equation}
z_l\sim U[0,1.5].
\end{equation}

Each system has an associated importance weight factor, $w_\mathrm{lens}$, which is defined as the probability to find a supernova in the whole sky area that reaches the limiting magnitude
\begin{equation}
w_{\mathrm{lens},{m}_{\mathrm{lim}}}= p(z_l) \int_{z_l}^{z_{s,\mathrm{max}}} \frac{dN_{\mathrm{SN}}}{dz_s}\frac{A_{\mathrm{SL}}}{A_{\mathrm{sky}}}(z_s,\sigma_l, z_l,q) P_{{m}_{\mathrm{lim}}}(z_s) .
\end{equation}

The probability distribution of lens redshifts $p(z_l)$ is calculated by normalising Equation \ref{eqn:pzl}. The lensing cross-section of the system is defined as $A_{\mathrm{SL}}$, and is then divided by the total arcseconds in the sky, $A_{\mathrm{sky}}$. The redshift-dependent rate of SNe~Ia is given by $dN_{\mathrm{SN}}/dz_s$ and $P_{{m}_{\mathrm{lim}}}$ is the probability of a supernova at some redshift to reach the characteristic detection depth of the survey. This is calculated as the fraction of supernovae in our simulations brighter than the limiting magnitude at peak. 
The sum of the weights factor is proportional to the estimated total discovery rate,
\begin{equation}
    R_{{m}_{\mathrm{lim}}}=f_{\mathrm{det},{m}_{\mathrm{lim}}}f_{\mathrm{sky}} \frac{N_{\mathrm{lenses}}}{N_{\mathrm{sim}}} \sum_{i=1}^{N_{\mathrm{sim}}} w_{i,{m}_{\mathrm{lim}}}
    \label{eqn:ratesim}
\end{equation}
where $f_{\mathrm{sky}}$ corresponds to the fractional area of the sky covered by the survey, $N_{\mathrm{lenses}}$ is the total number of galaxies in our lens redshift range, which is obtained by integrating Equation \ref{eqn:NLenses}, and $N_{\mathrm{sim}}$ is the number of lenses in our simulations. The value $f_{\mathrm{det},{m}_{\mathrm{lim}}}$ is the transient recovery fraction defined in Section ~\ref{sec:fdet}.

We carry out a simulation of lenses and sources to forecast rates and properties of multiply imaged SNe Ia for photometric surveys of different detection depths. To perform the simulation, we generate 100\,000 galaxies with parameters realized at random from the  distributions described in Section \ref{sec:simulations}. We then realize 1\,000 SNe positions at random in the area of the strong-lensing area for every $z_s \in \left(z_l\right.$,$\left.2\right]$. For each of these we simulate the unresolved supernovae light curves in the $i$- band for ZTF and LSST based on SNe~Ia spectral templates and find the $i$-band magnitude at peak for each of the light curves.

\section{Results and discussion}
\label{sec:results} 
For each of our limiting magnitudes, multiply imaged SNe Ia that are brighter than the limiting magnitude of the survey at peak are counted as "detectable" and their properties are stored. We first analyse the distributions for detectable systems for an idealized survey, where every glSNe brighter than our threshold would be detectable. The median values along with the $1\sigma$ deviations for these distributions can be found in Table \ref{tab:resultssummary}.

\begin{figure*}
\centering
\includegraphics[width=2\columnwidth]{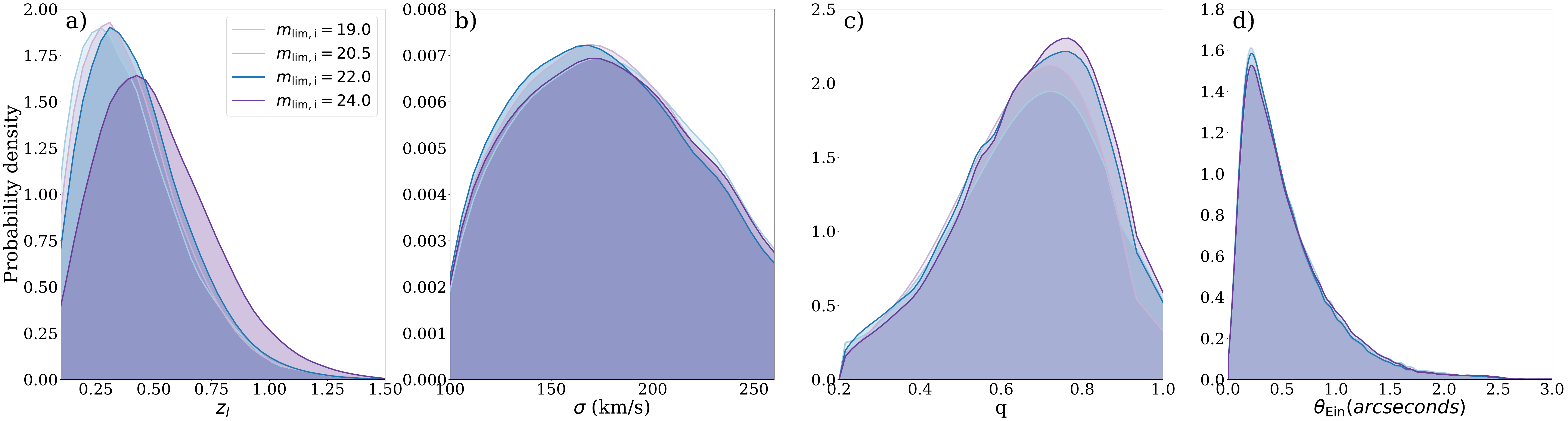}
\caption{\textbf{(a)} Distribution of lens redshifts. \textbf{(b)} Distribution of velocity dispersions. \textbf{(c)} Distribution of axis ratios. \textbf{(d)} Distribution of Einstein radii.}
\label{fig:lensdist2}
\centering
\end{figure*}

Figure \ref{fig:lensdist2} shows the properties of the lens galaxies for our detectable systems and the distribution of Einstein radii of the systems. The velocity dispersion and axis ratio distributions do not change significantly with the detection depth of the survey. Moving to deeper limiting magnitudes we are able to see sources at higher redshift, therefore it is expected that the distribution of lenses also moves towards higher redshifts where there is more volume and thus more potential lens galaxies. The Einstein radius distribution also remains constant. Since the velocity dispersion distribution stays constant, we conclude the unchanging Einstein radii distributions occur because of changes in the redshift distribution of both lenses and sources: lenses in deeper surveys are found at higher redshifts, but so are sources. 

\begin{figure*}
    \centering
    \includegraphics[width=2\columnwidth]{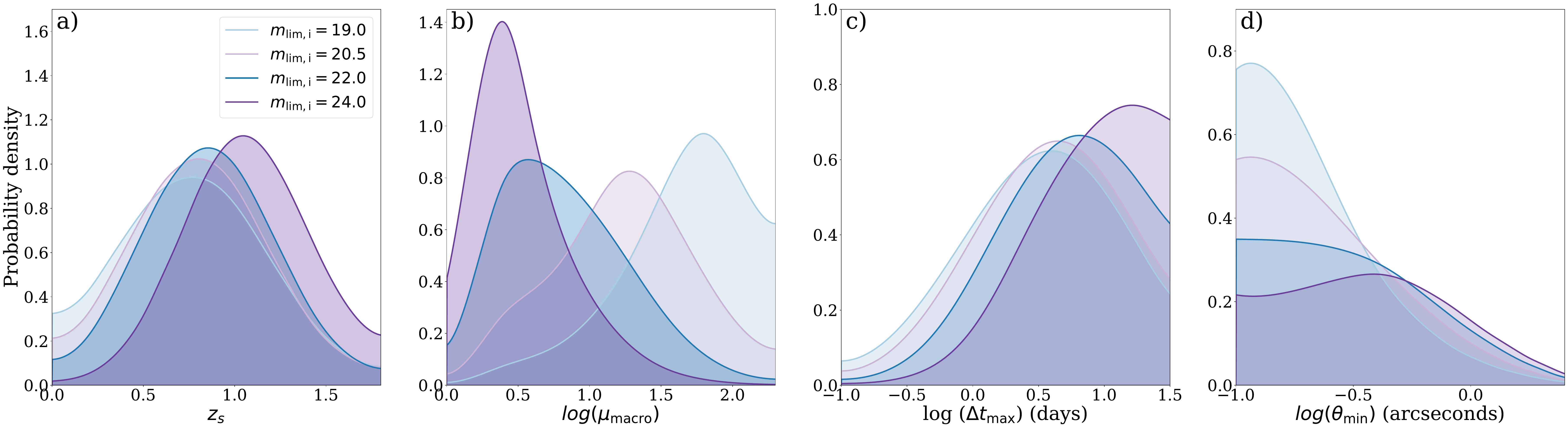}
    \caption{\textbf{(a)} Distribution of source redshifts. \textbf{(b)} Distribution of macromodel magnifications. \textbf{(c)} Distribution of time delays. \textbf{(d)} Distribution of minimum image separation.}
    \label{fig:mainparameters}
\end{figure*}

\begin{figure}
\centering
\includegraphics[width=\columnwidth]{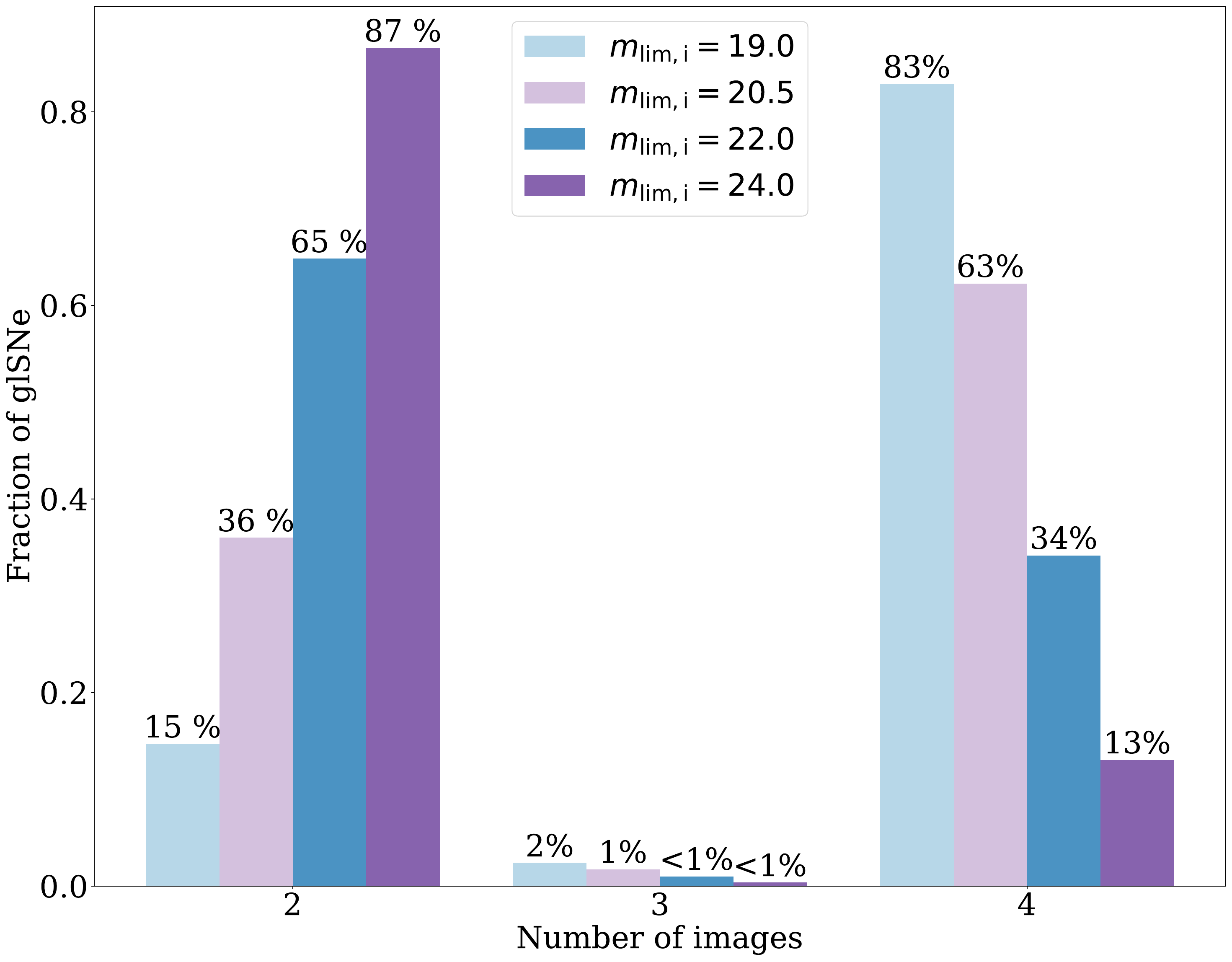}
\caption{Fraction of glSNe with two, three, and four images for the main magnitude thresholds considered in our discussion.}
\label{fig:nimages}
\centering
\end{figure}

Figures \ref{fig:mainparameters} and \ref{fig:nimages} show the distribution of properties of the sources and the galaxy-glSNe systems as a whole. We find that as we increase the detection magnitude of a survey, we detect fainter and hence higher redshift glSNe. Moreover, we find that the total magnification required to be detectable decreases significantly. Shallow surveys require magnifications in the order $\mu_{\mathrm{macro}}\approx 20$. Deeper surveys, in contrast, require total macromodel magnifications $\mu_{\mathrm{macro}}\approx 3$. We note that as depth of the survey increases, time delays get longer. This is unsurprising because sources are at higher redshift so time delay distances are larger. Moreover, the lower magnifications imply more asymmetric systems, which have longer time delays. The distribution of minimum angular diameter distance, or the resolution needed to fully resolve the system, is once again explained by the high magnifications required for detectability in shallow surveys. These high magnifications can only be achieved if the source is located close to the caustics. This leads to images forming close together and appearing with short time-delays. As we move to fainter detection thresholds, we do not require the source to be located close to the caustics, so we find image configurations that are located further apart. 

Regarding the multiplicity of the systems, we see an evolution in the double-to-quad ratio with survey detection depth. Shallow surveys are dominated by quads, once again an effect of the need for higher magnifications to be detectable in these surveys. We also find a small number of systems with three-image configurations. These are naked-cusp configurations \citep{nakedcusp} formed by lenses with large ellipticity. Since these are intrinsically high magnification configurations it is not surprising that we find the naked-cusp fraction decreases as the survey depth increases. We find mostly two-image systems in deeper surveys.

Since the minimum total magnification for strong lensing is 2 for an SIE lens and the median magnification from our simulations is $\mu_{\textrm{macro}} \approx 3$ (Fig.~\ref{fig:mainparameters}(b)), this implies that essentially all strongly lensed SNe peak with a magnitude brighter than $m_i<24$. We draw a main conclusion from these distributions: shallow surveys are magnification selected, whilst deep surveys are limited by the volume of the Universe that is strongly lensed. 

This is consistent with the analytical formalism presented by \cite{oguri--19} and can be attributed to the slope of the lensing rate as a function of redshift. For lower redshift sources, which are predominant in more shallow surveys, events with $\mu>>1$ are preferred, which explains the magnification factors of iPTF16geu and SN~Zwicky. For higher limiting magnitudes, the slope becomes more shallow and highly magnified events are not necessarily preferred.
  
\begin{table*}
 \caption{Median values describing the population of glSNe detectable in the i-band}
 \label{tab:resultssummary}
 \begin{tabular}{llllll}
  \hline
  Parameter &Description & $m_{\textrm{lim,i}}=19$ mag & $m_{\textrm{lim,i}}=20.5 $ mag & $m_{\textrm{lim,i}}=22$ mag & $m_{\textrm{lim,i}}=24$ mag\\
  \hline
  $z_l$ & Lens redshift & $0.35_{-0.18}^{+0.26}$ & $0.37_{-0.18}^{+0.25}$& $0.39_{-0.19}^{+0.25}$ & $0.48_{-0.21}^{+0.28}$\\
  $\sigma$ & Lens velocity dispersion (km s$^{-1}$) &  $185_{-51}^{+61}$ & $182_{-50}^{+60}$ & $180_{-49}^{+62}$ & $184_{-51}^{+63}$\\
  $q$ &  Lens axis ratio &  $0.71_{-0.19}^{+0.12}$ & $0.68_{-0.19}^{+0.15}$ & $0.68_{-0.2}^{+0.16}$ & $0.7_{-0.2}^{+0.15}$\\
  $z_s$ &  Source redshift & $1.05_{-0.23}^{+0.27}$ &$0.82_{-0.33}^{+0.28}$ & $0.87_{-0.32}^{+0.26}$ &$1.06_{-0.23}^{+0.28}$ \\
$\mu_{\mathrm{macro}}$& Magnification due to macromodel & $50_{-3}^{+2}$ & $17_{-3}^{+3}$ & $6_{-3}^{+3}$ & $3_{-2}^{+3}$\\
 $ \Delta t_{\mathrm{max}}$& Maximum time delay (days) & $4_{-4}^{+4}$ & $4_{-4}^{+4}$ & $6_{-2}^{+3}$ & $10_{-4}^{+3}$\\ 
 $\theta_{\mathrm{max}}$ &Separation between images with max. $\Delta t$ (arcseconds)& $0.82_{-0.48}^{+0.92}$ & $0.80_{-0.47}^{+0.88}$ & $0.79_{-0.36}^{+0.90}$ & $0.83_{-0.48}^{+0.96}$\\
  $\theta_{\mathrm{Ein}}$ & Einstein radius (arcseconds)& $0.44_{-0.25}^{+0.48}$ &$0.42_{-0.24}^{+0.46}$ & $0.41_{-0.23}^{+0.47}$ & $0.43_{-0.24}^{+0.49}$  \\
$\theta_{\mathrm{min}}$ &Minimum separation between images (arcseconds)& $0.06_{-0.05}^{+0.39}$ & $0.25_{-0.25}^{+0.62}$ & $0.48_{-0.26}^{+0.67}$ &$0.65_{-0.41}^{+0.66}$\\
  \hline
 \end{tabular}
\end{table*}

\subsection{iPTF16geu and SN~Zwicky in context}
\label{sec:comparison}

Having studied the properties of detectable systems for different photometric surveys in detail, we can now compare these distributions to the two glSNe~Ia found to date: iPTF16geu and SN~Zwicky. Both shared some similar properties: they were highly magnified quads found in systems with low angular separation and with time delays of $\lesssim1$ day that did not allow for precise time delay measurements. iPTF16geu was detected by the intermediate Palomar Transient Factory, which used the same telescope as ZTF to observe the sky to a similar depth, but with lower cadence and a smaller area. Both iPTF16geu and SN~Zwicky were discovered serendipitously. Spectra were taken as part of systematic followup. iPTF16geu  was seen to be $4.3$ mag brighter than a typical SN~Ia at the measured redshift \citep{goobar--17} and SN~Zwicky was $3.5$ magnitudes brighter \cite{goobar--22}. Thus, we compare their properties with the population of glSNe detectable at $m_{\textrm{lim}}=19$ mag, which corresponds approximately to the cutoff for the ZTF Bright Transient Survey \citep{fremling--20} which provided nearly complete spectroscopic followup of bright ZTF transients.

\begin{figure}
    \centering
    \includegraphics[width=\columnwidth]{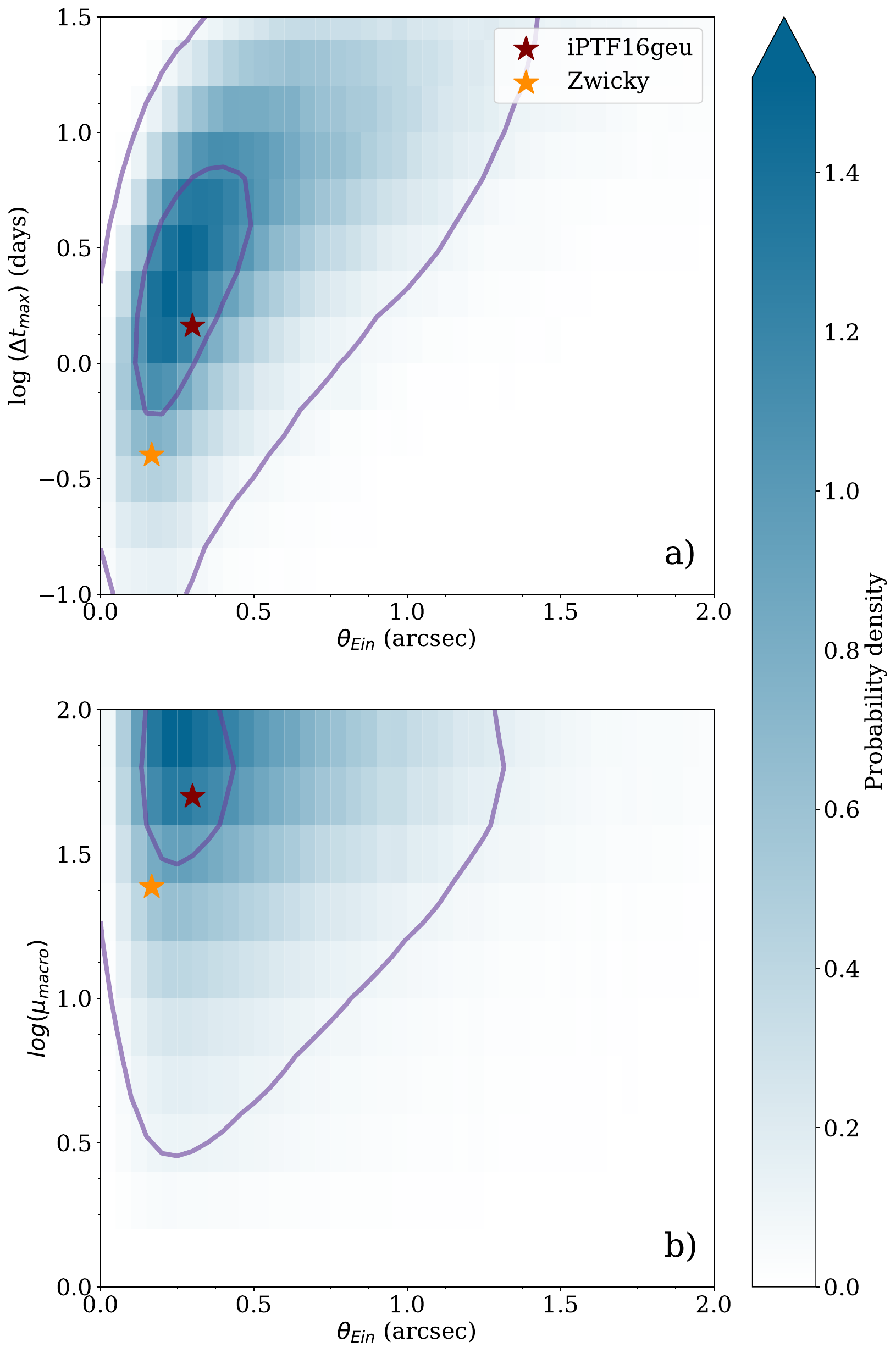}
    \caption{Distributions of \textbf{(a)} Einstein radius and maximum time delays and \textbf{(b)} Einstein radius and total macromodel magnification for a survey with $m_{\textrm{lim,i}}=19$ mag. The 1 and 2$\sigma$ contours are delineated by the purple lines. The orange and red stars represent iPTF16geu and SN~Zwicky in this parameter space. Both events are found within the $2\sigma$ contours.}
    \label{fig:zwicky}
\end{figure}

In Figure \ref{fig:zwicky}, we show the distribution of time delays and total magnification as a function of Einstein radius for lensed transients peaking brighter than $m_i = 19$~mag. iPTF16geu is within the 68 percent likely region in this parameter space, whilst  SN~Zwicky is within the 95 percent contour. We therefore conclude that iPTF16geu is typical of what we should expect for glSNe peaking above $m_i = 19$~mag. The lens in SN Zwicky is somewhat lower mass than is typical, but there is no indication that SN Zwicky is a substantial outlier.

\begin{figure}
    \centering
    \includegraphics[width=\columnwidth]{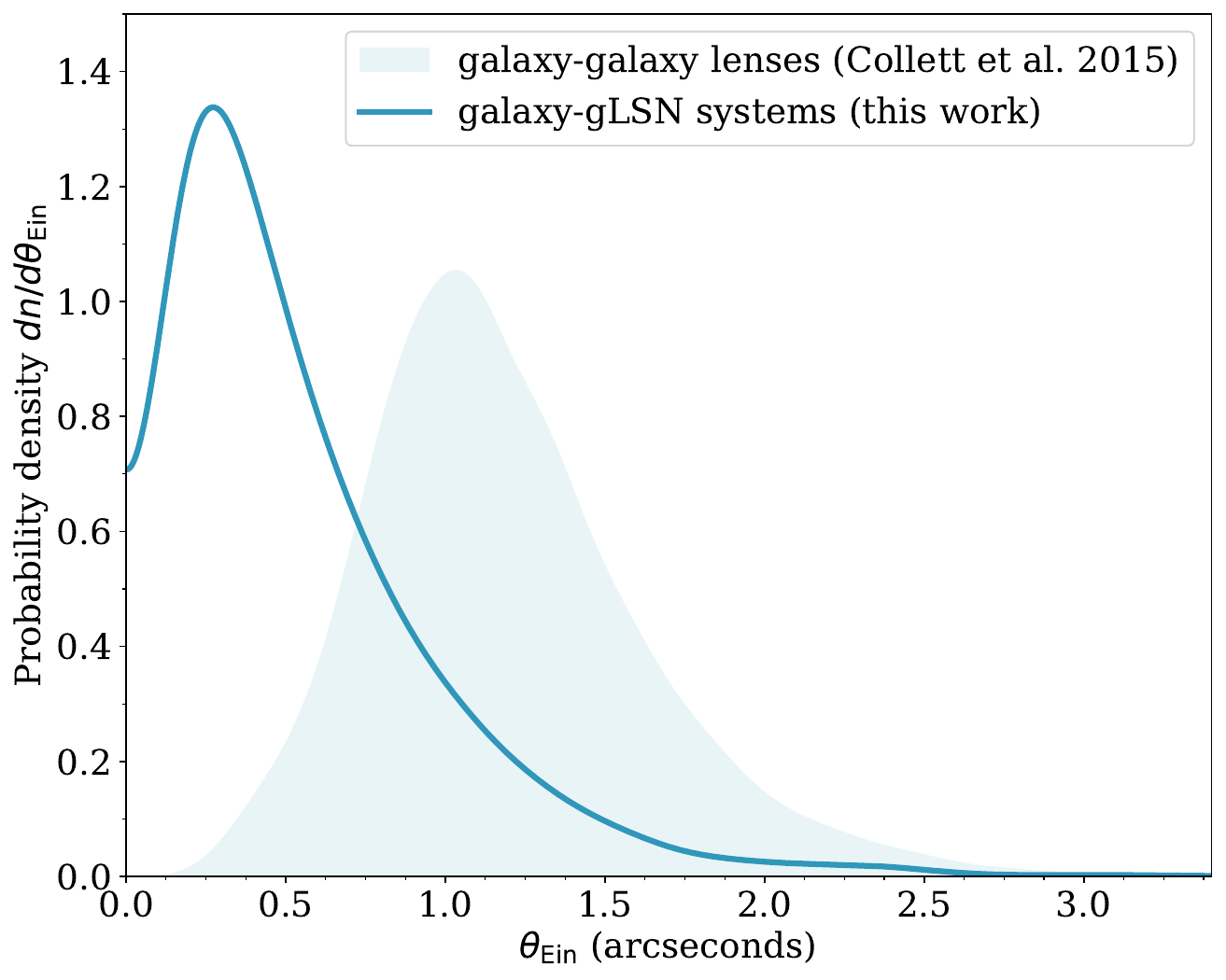}
    \caption{Distribution of Einstein radii for galaxy-glSNe lensing systems for a survey with $m_{\rm{lim,i}}=24.0$ mag compared with the distribution of Einstein radii for galaxy-galaxy lens systems for LSST from \citealt{Collett_2015}.}
    \label{fig:lenspop}
\end{figure}

Whilst iPTF16geu and SN~Zwicky are consistent with what we should expect of shallow surveys, this does not mean that they are representative of what we will find in upcoming deeper surveys. Given the distribution shown in Fig.~\ref{fig:mainparameters}, as detection depth becomes fainter events like these will become outliers. The population of glSNe~Ia will be dominated by systems with longer time delays and lower magnifications. An exception is the small Einstein radius of the lenses, which appears to not be a unique property of these two systems and does not change with the detection depth of the survey. We compare the predicted Einstein radius distribution for detectable galaxy-glSNe systems with the predictions for galaxy-galaxy lenses detectable by LSST \citep{Collett_2015}, and conclude that glSNe~Ia are found in much more compact image configurations, i.e. with smaller Einstein radii than galaxy-galaxy systems, as shown in Fig.~\ref{fig:lenspop}. The median Einstein radius for galaxy-glSN systems in a survey like LSST is 0.36\farcs{}, whereas for galaxy-galaxy lenses the median Einstein radius will be 1.15\farcs{}. Further comments on the implications of this in the context of glSNe~Ia discoveries are found in Section ~\ref{sec:discovery}, but this shows that a large fraction of glSNe~Ia in LSST will not be located in known galaxy-galaxy lenses. Moreover, this result highlights the importance of glSNe~Ia to study and properly characterise low mass lenses. The discovery of new glSNe will give us insights into a new population of faint and compact lensing systems under-represented in galaxy-galaxy lensing systems \citep{goobar--22}.

\subsection{glSNe~Ia rate forecasts and prospects for cosmography}

One of the main questions in the field of glSNe~Ia is how many systems are going to be detected in the era of wide-field time-domain surveys. As we discuss in Section \ref{sec:weights}, our simulations also provide an estimate of yearly glSNe~Ia yields. In practice, not all the glSNe that reach the detection threshold will be observed. We follow the methodology in Section ~\ref{sec:fdet} to define our observing strategy and determine what fraction of glSNe~Ia brighter than our survey's limiting magnitude will be detected.  We can then use these results to estimate the yields of glSNe~Ia we expect to discover with this method. 

Using our ZTF-like survey plan, we estimate to detect 17\% of glSNe that have an $i$-band peak magnitude brighter than 20.5~mag when requiring two 5$\sigma$ detections. Together with Equation \ref{eqn:ratesim}, this implies we should discover $\approx 0.2$ glSNe~Ia at $m_{\textrm{lim,i}}=20.5$ per year with ZTF.  Most of these glSNe were initially detected in the $r-$band. This is due to the higher cadence and coverage of the survey in this filter. In the interval $z_s \in \left[1.3, 1.5\right]$, however, all glSNe~Ia were discovered in the $i$-band, showing the importance of redder filters for glSNe discovery at higher redshifts. As expected, no glSNe were discovered beyond $z=1.5$.This is consistent with the fact that we have detected one glSN in the five years of the survey. Changing our detectability requirement to three or four detections at the $5\sigma$-level does not have a significant impact on yields, with the requirement for four detections allowing us to recover 15\% of our sample. This is due to the high cadence of ZTF. Setting the requirement of $m_{\textrm{lim,i}}=19$ mag, we estimate there is only a 1 in 50 chance of finding a glSN bright enough to trigger spectroscopic classification through the BTS. These rates imply that iPTF16geu and SN~Zwicky are fortuitous events. Following a similar method with the LSST observations survey plan, we find that we should be able to recover $\approx 16$\% of glSNe when requiring two 5$\sigma$ detections. We expect 18 glSNe~Ia per year with LSST, increasing the order of magnitude of our sample by a factor of $\mathcal{O}$($10^2$) over the 10 year LSST survey. 

\begin{figure}
    \centering
    \includegraphics[width=\columnwidth]{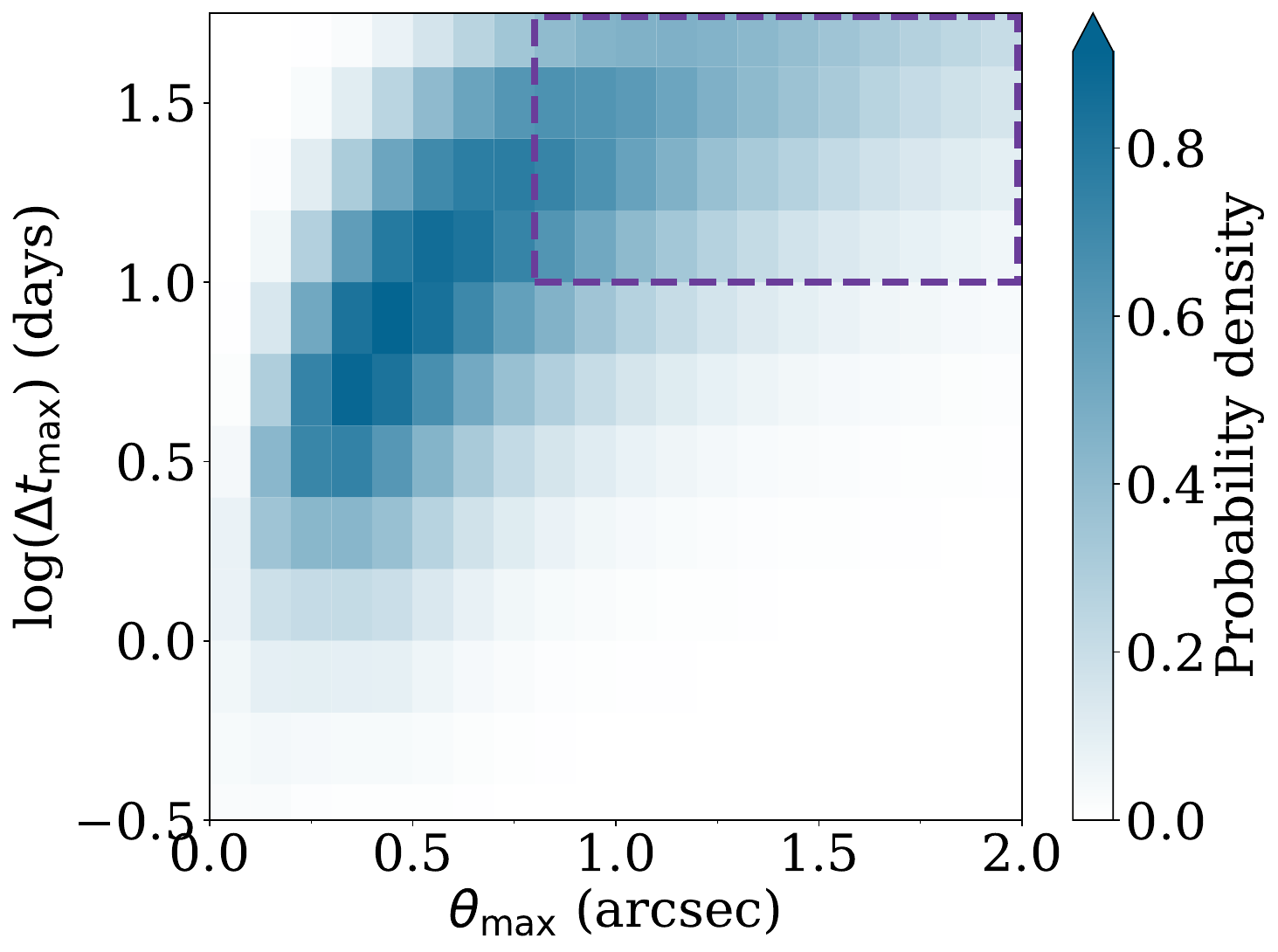}
    \caption{Maximum time delay vs. maximum separation distribution for $m_{\textrm{lim,i}}=24$ mag. The dashed lines enclose the region of parameter space containing glSNe that we defined as "cosmologically useful". From our simulations, 38\% of glSNe~Ia fall within this region.}
    \label{fig:cosmology}
\end{figure}

Given this increase in the number of glSNe~Ia discovered, one might ask what fraction of these will be suitable for time delay measurements. In order to attempt to answer this question we must first define what makes a system "cosmologically useful". We define a system with a "good" time delay measurement as one with time delay $\Delta t_\mathrm{max}\geq 10$ days. Previous work has shown that time-delay uncertainties are likely to be at the 1 day level \citep{holismokestd,piereltd}. Given a 10-day delay this translates to a 10\% precision in the time delay measurement and therefore a 10\% precision in our $H_0$ measurement assuming a good understanding of the lens model. For a survey with the limiting magnitude of LSST, $50\%$ of glSNe will pass this cut, as seen in Table \ref{tab:resultssummary}. For a a good glSN system, we also require that the two images with the longest time delay are resolvable. Despite there being attempts to obtain time delays from unresolved glSNe~Ia light curves \citep{unresolvedlc}, these do not take into account the effects of microlensing. \cite{Goldstein_2018} showed that microlensing can introduce time-delay uncertainties of $\sim 4\%$ into the light curves of typical LSST glSNe~Ia. This number could be reduced to $1\%$ if measuring time-delays from achromatic-phase colour curves instead of light curves. Measuring reliable colour curves requires the images to be resolved, so we set the condition for the brightest images to be separated by an angular distance $\theta_{\mathrm{max}}\geq0.8''$, using the 75th percentile of the effective FWHM \citep{lsstsciencebook}. Figure \ref{fig:cosmology} shows that $38\%$ of glSNe in LSST are found in this region of parameter space. Follow-up with a space telescope would remove this requirement, but it is unlikely that sufficient time will be available to do cadenced monitoring of many glSNe from space.

Combining both the time-delay and resolution requirements implies that we should discover $\approx 7$ systems per year that are suitable for time-delay cosmography. By the end of observations with LSST, we should have $\approx 68$ "cosmologically useful" glSNe~Ia with a time delay measurement with 10\% precision. Assuming other systematic errors can be kept under control, this would lead to a value of $H_0$ from glSNe~Ia to 1.2$\%$ precision in the next 10 years.

In reality, due to the low cadence of LSST, it is unlikely that this survey alone can give us good enough sampling of the light curves to obtain precise time delay measurements. \cite{huber19} point out the importance of follow-up observations, proposing the use of LSST as a discovery tool for glSNe~Ia. They show that for time delays of at least 20 days, a strategy combining LSST and follow-up observations allow for time delay measurements with a 1\% error. In our simulations, 22\% of the systems have $\Delta t_{\mathrm{max}}\geq$ 25 days and image separations $\theta_{\mathrm{max}}\geq0.8''$, yielding $\approx$ 4 glSNe~Ia per year with these characteristics. The downside of these high precision time-delay measurements is that other systematic errors start to become dominant. These results highlight the great prospects for glSNe~Ia cosmology in the upcoming LSST era.

\subsection{Comparison with previous analyses} 

The properties of glSNe in our simulations are consistent with previous studies \citep{marshall_2010, quimby--14, More_2017, goldstein--19, wojtak19}. Some differences arise from the fact that our simulations aim to study the properties of glSNe that are bright enough to be detected. By using only brightness as a selection criterion, without relying on additional glSNe identification strategies, we aim to reduce selection biases in the properties of our sample. For example, the method used by \cite{marshall_2010} includes resolved light curves, constraints on the image separation and flux ratios of the images, and requires the faintest image in a double system or the third brightest image in a quad system to be detectable. Many of our glSNe would not have passed these detection cuts, as most glSNe in ZTF and LSST will not be resolvable. On the other hand, \cite{quimby--14} use a colour-based selection method, and therefore account for detection from only the brightest image. The aim of their work is to understand the properties of PS1-10afx, the first glSN discovered. Due to this, their simulations fix ellipticity and lens and source redshifts to those of PS1-10afx. \cite{goldstein--19} carry out pixel-level simulations of glSNe accounting for the observing strategy of LSST and follow the discovery technique described by \cite{Goldstein_2018}, which is based on photometric classification through spectral template fitting. \cite{wojtak19} consider two discovery techniques based on image multiplicity and high magnification, as well as a hybrid method combining these two.

The discovery technique will also have an impact on rate estimates. Our predicted yields are consistent with those presented by \citep{goldstein--19} and \cite{wojtak19} within an order of magnitude. \cite{goldstein--19} predict $\approx$ 1 glSN~Ia per year discovered by ZTF and $\approx$ 47 glSNe~Ia for two different observing strategies in LSST. For ZTF however, \cite{goldstein--19} assumed that one week of data would be stacked for detection, so it is not surprising that the \cite{goldstein--19} ZTF forecast rates are a factor of five higher than ours. \cite{wojtak19} predict ZTF should detect $\approx$ 2 glSNe~Ia per year, while LSST rates are $\approx 90$ for their most efficient method. 
\cite{magee2023} analyse four years of observations from the ZTF public survey and estimate $\lesssim1$ glSN during this time using a similar colour-based method to \cite{quimby--14}, showing consistency with our predictions.

As we will discuss in Section ~\ref{sec:assumptions}, accurate rate estimates depend on a large number of assumptions. Both of the analyses in \cite{goldstein--19} and \cite{wojtak19} were carried out with previous versions of the LSST observing strategy, which under-accounted for overheads. We also highlight that our analysis includes realistic lightcurves which account for the effects of microlensing. In reality, many images will be demagnified as a result of microlensing from stars in the lens plane, pushing them beyond detectability. Other systems will be more magnified, requiring lower macromodel magnifications in order to be detectable. As a consequence, microlensing increases the double-to-quad ratio for detectable systems, with the largest effect seen in shallow surveys. The scatter due to microlensing for each macro-magnification in our analysis is shown in Figure \ref{fig:microlensingscatter}. In Figure \ref{fig:mlrates}, we compare simulations in the $i$-band with and without microlensing and demonstrate the impact microlensing has on the glSNe yields. For shallow surveys, the rate of detectable glSNe increases when accounting for the effects of microlensing, as the microlensing magnification distribution pushes some glSNe above the detection threshold, while those that are demagnified were unlikely to be detectable even before these effects. For a survey with the detection depth of ZTF, microlensing increases the detected glSNe rate by 15\%, increasing the chances of discovering systems such as iPTF16geu or SN~Zwicky. For surveys with limiting magnitude $m_{\textrm{lim,i}} \geq 22.5$, the balance reverses, with microlensing pushing more systems below the detection threshold than it brings into discoverability. For a limiting magnitude of $m_{\textrm{lim,i}} = 24$, 8\% of systems are lost due to microlensing. We conclude that we will lose some potential discoverable systems due to microlensing in upcoming deep surveys, but the main uncertainties in rate estimates will still derive from the survey plan and lens modelling.

\section{Limitations}
\label{sec:discussion} 
As previously mentioned, rate forecasts depend on a wide range of assumptions and should therefore be taken as a general estimate. In the following section, we comment on the limitations of our work that have an impact on the rates of glSNe~Ia detectable by the different surveys. Our simulations depend on our knowledge about the deflector and source population and our detectability conditions for each survey. Despite being well-motivated by observations, these models are based on a series of assumptions that could bias our predictions. We also discuss the challenges that come with discovering glSNe~Ia and what the approach should be to overcome these challenges.
\subsection{Assumptions in our simulations}
\label{sec:assumptions} 
We first comment on the limitations of our lens and source models. We assumed the velocity dispersion does not evolve with redshift. For lower redshifts, this is consistent with results from \cite{vdf1} and \cite{Shu_2012}, who find little variation in the velocity dispersion function up to redshift $z\approx 0.5$. Both studies find an increase in variability at higher redshifts. As most of the lenses in our systems are found at redshifts $z \lesssim 0.5$ however, this assumption is likely appropriate.

For simplicity, we assumed all of our lenses follow an SIE density profile. Whilst this has been shown to be a good approximation for massive eliptical galaxies \citep{auger_2009}, it is unknown for lower mass lenses. Assuming a different lens model with a different density slope will impact the distribution of time delays and magnifications for each of the images, affecting our predicted rates.

Further work could be done to quantify the impact of the lens model on rates. Our simulations do account for the effects of microlensing by stars in the lens galaxy. A more detailed discussion regarding the impact microlensing has on rates is given in Appendix \ref{sec:MLrates}.

We also made some simplifications in our glSNe simulations. We made the choice of the \cite{hsiao--07} templates because they extend to shorter wavelengths than other templates, which allowed us to model glSNe at higher redshifts. We accounted for scatter in the bolometric magnitude of unlensed SNe, but we did not introduce known correlations between other SNe~Ia parameters, e.g. stretch, colour \citep{Hoeflich_2017,sne2020,environmentaleffects}. While we do not account for extinction when calculating the number of detectable glSNe, our \texttt{simsurvey} simulations do include host galaxy and Milky Way extinction. We follow the methods of \cite{feindt--19} to model extinction due to the host by drawing from a random exponential with a rate $\lambda=0.11$. We did not account for extinction due to dust in the lens galaxy, which will reduce the rates of detectable glSNe by making them appear fainter. Finally, we did not include contamination from the light of the host and the lens, which would make glSNe more difficult to detect. These assumptions imply our rates are more likely to be an upper limit.

We also made some simplifications when defining our survey plan. For the case of ZTF, we used the survey plan from \cite{feindt--19} with a constant $m_{\textrm{lim}}=20.6$ mag, not accounting for other factors such as weather. The rate of detectable glSNe will be highly dependent on our survey strategy. Previous work (e.g. \citealt{huber19}) has shown the impact of observing cadence for different LSST strategies will have on glSNe~Ia discovery and time delay measurements. The current observing strategy of LSST is based on a rolling cadence in the WFD survey which is not ideal for glSN discovery \citep{huber19}. Whilst we have not quantified the impact of these assumptions, none are likely to introduce substantial errors on our rate forecasts, but we stress these should be taken as an upper limit.

\subsection{The challenge of glSNe discovery}
\label{sec:discovery}

We have shown that LSST has the potential to discover over a hundred systems in its 10-year survey, which could lead to a measurement of $H_0$ with $\approx$ 1\% precision. Identifying these systems as detectable however does not mean they will necessarily be discovered. The small sample size of glSNe can be explained in part by the difficulty of detecting them.

Several methods have been proposed to search for glSNe~Ia. Perhaps the most straightforward one consists of monitoring known lenses \citep{craig--21}. As we showed in Section \ref{sec:comparison}, other methods are needed to complement this one, as most glSNe will be found in systems that cannot be detected in galaxy-galaxy lens searches. Following from the first ever detected glSN, PS10-afx, \cite{quimby--14} suggested a method based on looking for transients that appear extremely red based on their magnitude. \cite{Goldstein_2018} proposed a method based on spectral template fitting to flag objects that are inconsistent with the light curve of an unlensed SNe~Ia at the apparent host galaxy redshift, and claimed that a method for reducing the number false positives is to only look at SNe in ellipticals. \cite{wojtak19} considers two discovery techniques based on image multiplicity and high magnifications, the latter being similar to the one proposed in this work, and found that for deeper surveys the image multiplicity method will be more efficient at finding glSNe.

\cite{magee2023} tested several of these methods on real data from the ZTF public survey and conclude none of these would have worked in the detection of SN~Zwicky. They find the false positive rate in this methods makes them infeasible in the upcoming era of wide-field surveys, as the alert pipeline would be flooded with false positives.

\section{Conclusions}
\label{sec:conclusions} 
In this work, we have shown how the properties of the population of detectable glSNe~Ia evolve with the survey's limiting magnitude through detailed and updated simulations including the effects of microlensing and the most up-to-date LSST observing strategy. This allowed us to put iPTF16geu and SN~Zwicky into context, make forecasts for glSNe~Ia rates in upcoming surveys, and comment on prospects for precise time-delay measurements in the LSST era.

We have built a realistic model of elliptical lens galaxies and lensed SNe~Ia to simulate a population of galaxy-glSN systems representative of those in the Universe. We then studied how the properties of these systems vary with a survey's detection depth, and found that the properties of the deflector population remain mostly unchanged, aside from the lens redshift. We found that as surveys go deeper, they will discover more lenses at higher redshift. This is explained by the fact that these surveys will observe a larger volume of the Universe, where there are more galaxies.

We have found that the properties of shallow surveys are defined by the high magnifications required to detect glSNe~Ia, whereas deep surveys are limited by the volume of the Universe that is strongly lensed. As a result, the population of glSNe~Ia in shallow surveys is dominated by highly-magnified systems of four images with short time-delays and low image separations. These correspond to sources located close to the caustics. We showed that as the survey's limiting magnitude increases, the distribution of glSNe~Ia properties begins to resemble the initial distributions that went into our simulations. At a limiting magnitude $m_{\mathrm{lim,i}}=24$, the dominant population consists of more asymmetric two-image systems, with lower magnifications, longer time delays and greater angular separations. These distributions allowed us to show that iPTF16geu and SN~Zwicky were both found within the 95\% confidence interval of the parameters that would be expected for a very shallow survey with $m_{\mathrm{lim,i}}=19$ mag.

By comparing the Einstein radii of galaxy-glSNe systems with forecast galaxy-galaxy populations, we showed that a large fraction of glSNe~Ia will not be found in known galaxy-galaxy lenses. Thus, even if monitoring known lenses can be used as a method to discover glSNe~Ia \citep{craig--21}, focusing solely on this method will not be sufficient. On the other hand, our forecasts show that glSNe~Ia will allow us to study low mass lenses that cannot be discovered through galaxy-galaxy lensing.

Our simulations allowed us to obtain forecast rates of glSNe~Ia that could be discovered by current and future photometric surveys. In order to do this we built survey plans for ZTF and LSST and estimated the fraction of glSNe in our simulations that would be detected when requiring two 5$\sigma$ detections. We found that a survey with the characteristics of ZTF should discover 1 glSN~Ia in a five-year survey. We also predicted that a ten-year survey based on the current baseline observing strategy of LSST will be able to detect $\approx 180$ glSNe~Ia systems. These forecasts broadly agree with previous rate estimates \citep{Goldstein_2018, goldstein--19, wojtak19} on the fact that wide-field surveys will increase our sample of glSNe~Ia by $\mathcal{O}$($10^2$), allowing us to produce the first statistical samples. The precision of these estimates however, is highly dependent on other factors such as the discovery technique and the survey observing strategy.

We showed that $\approx 38\%$ glSNe~Ia found in LSST will allow for time-delay measurements within 10\% precision. Based on our previous rate estimates, by the end of observations with LSST we should find up to $\approx70$ glSNe~Ia with precise time-delay measurements, leading to a $\approx$ 1.2$\%$ estimate of $H_{\mathrm{0}}$. 

In this work we have focused on SNe~Ia due to their standardisable nature that can break the mass sheet degeneracy, and their well-understood light curves allowing for easy time delay measurement. Our simulations could be extended to study the properties of other glSNe types, including type II SNe. These SNe pose other challenges such as the greater variety of spectral templates, their generally longer lifetime, and their light curve shapes that make time delay measurements from photometric data less straightforward. The forecasts in \cite{goldstein--19} show more glSNe of these types can be found per year with both ZTF and LSST. This, along with the fact that their spectra can be used to measure time-delays with $\leq 2$ day uncertainty \citep{dttype2}, points towards the importance of extending the area of glSNe cosmography beyond SNe~Ia. 

Whilst iPTF16geu and SN~Zwicky have not proven to be cosmologically useful, this work highlights that they are not representative of the glSNe that will be discovered in deeper surveys. The field of glSNe cosmography should soon yield precise constraints on the Hubble constant, but in order to exploit these opportunities our efforts should now focus on finding the most efficient methods to promptly identify these systems and solve the challenge of false positives.

\section*{Acknowledgements}
We are grateful to Nikki Arendse, Suhail Dhawan, Justin Pierel, Dan Ryczanowski, Ana Sagués Carracedo, Ariel Goobar and Paul Schechter for helpful conversations that have enriched this work.

Numerical computations were done on the Sciama High Performance Compute (HPC) cluster which is supported by the ICG, SEPNet, and the University of Portsmouth.

This work has received funding from the European Research Council (ERC) under the European Union's Horizon 2020 research and innovation programme (LensEra: grant agreement No 945536). TC is funded by the Royal Society through a University Research Fellowship. MRM acknowledges a Warwick Astrophysics prize post-doctoral fellowship made possible thanks to a generous philanthropic donation.
For the purpose of open access, the authors have applied a Creative Commons Attribution (CC BY) license to any Author Accepted Manuscript version arising.

%%%%%%%%%%%%%%%%%%%%%%%%%%%%%%%%%%%%%%%%%%%%%%%%%%
\section*{Data Availability}
All data are simulated: our simulations are available from the corresponding author upon request.



\bibliographystyle{mnras}
\bibliography{lensedSN} 
%%%%%%%%%%%%%%%%%%%%%%%%%%%%%%%%%%%%%%%%%%%%%%%%%%

%%%%%%%%%%%%%%%%% APPENDICES %%%%%%%%%%%%%%%%%%%%%

\appendix

\section{Additional figures}
\label{sec:MLrates}
\begin{figure}
    \centering
    \includegraphics[width=\columnwidth]{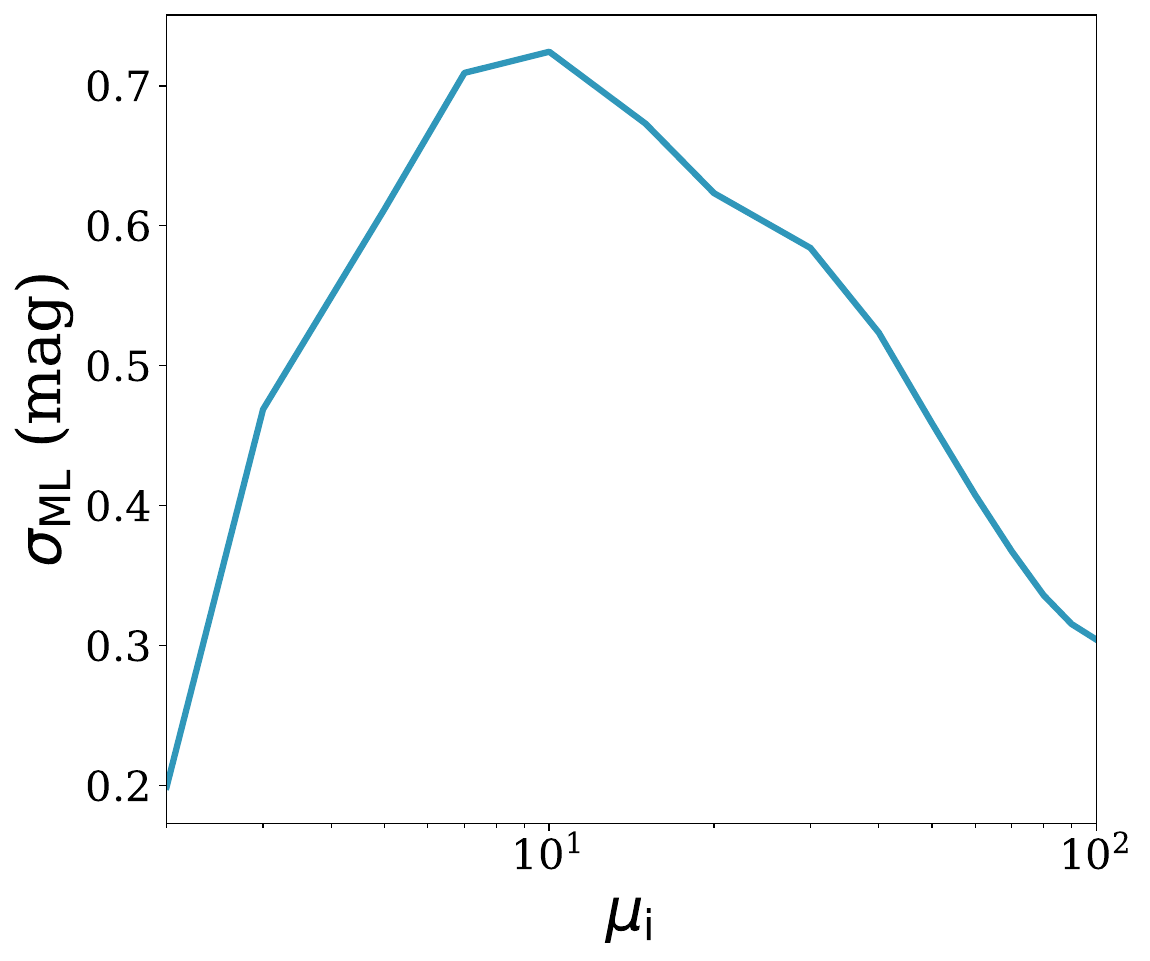}
    \caption{Scatter introduced due to microlensing for a lens with stellar fraction $s_*=0.5$. We find the peak in microlensing scatter for an image with macromodel magnification $\mu_{\mathrm{i}}=10$.}
    \label{fig:microlensingscatter}
\end{figure}

\begin{figure}
    \centering
    \includegraphics[width=\columnwidth]{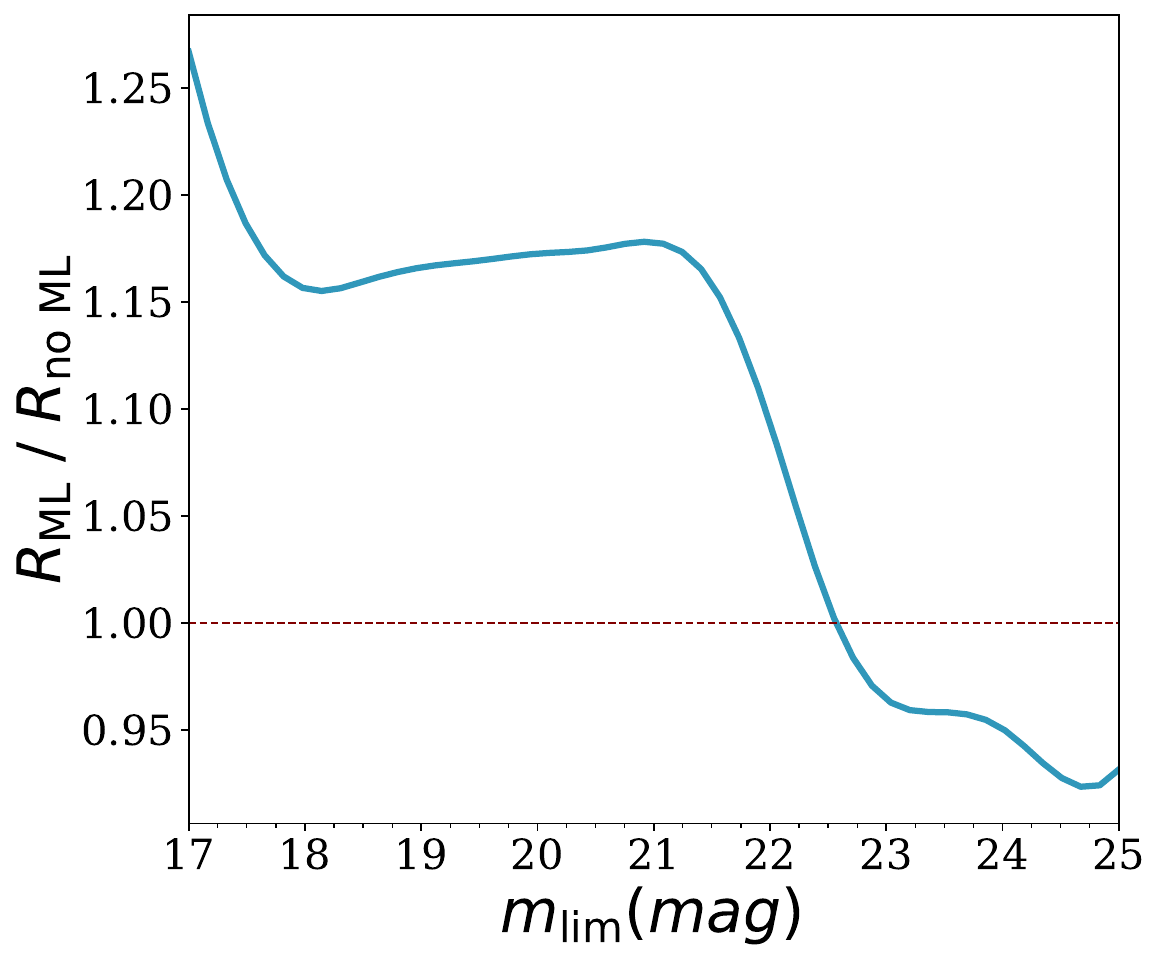}
    \caption{Ratio of SNe per year discoverable when accounting for microlensing ($R_\mathrm{ML}$) and not accounting for microlensing ($R_{\mathrm{no \ ML}}$). The dashed line represents the value $R_{ML}= R_{\mathrm{no \ ML}}$.}
    \label{fig:mlrates}
\end{figure}

%%%%%%%%%%%%%%%%%%%%%%%%%%%%%%%%%%%%%%%%%%%%%%%%%%

% Don't change these lines
\bsp	% typesetting comment
\label{lastpage}
\end{document}